\documentstyle[12pt,mont98_desy,amssymb,epsfig]{article}

\def\be{\begin{equation}}
\def\ee{\end{equation}}
\def\bea{\begin{eqnarray}}
\def\eea{\end{eqnarray}}

\begin{document}

\title{II.~~LONGITUDINAL ELECTRON SPIN POLARISATION AT 27.5 GEV IN HERA 
\footnote{A talk presented 
at the 15th ICFA Advanced Beam Dynamics Workshop: ``Quantum
Aspects of Beam Physics'', Monterey, California, U.S.A., January 1998.
Also in DESY Report 98--096, September 1998.}
}

\author{  D.P.~BARBER \\{\it for the HERA Polarisation Group. } }

\address{Deutsches Elektronen--Synchrotron, DESY, \\
 22603 Hamburg, Germany. \\E-mail: mpybar@mail.desy.de}

\maketitle
\abstracts{The first attainment of longitudinal spin polarisation
in a high energy electron (positron) storage ring is described.}

\section*{Summary}
An integral part of the design of the HERA $ep$ collider 
has been the provision of longitudinally spin polarised electrons for the high
energy physics experiments at the interaction points \cite{ba95}.
At HERA, electrons or positrons of energy up to $30~GeV$ are brought into
collision with $820~GeV$ protons.

As outlined in Article I,  stored electron beams can become vertically
polarised by the Sokolov--Ternov effect \cite{st64}. The maximum ST 
polarisation
achievable is $92.4\%$ corresponding to a planar ring.
              To provide longitudinal polarisation at an interaction
point the naturally occuring vertical polarisation in the arcs must be
rotated into the longitudinal direction just before the interaction
point (IP) and back to the vertical just after the IP using special
magnet configurations called spin rotators.

At HERA the spin rotators consist of strings of interleaved horizontal
and vertical bending magnets each of which deflects the orbit by no more
than about $20$ milliradians. Dipole rotators exploit the prediction of 
Eq.~(4) in Article I that for motion transverse to the magnetic field,
the spin precesses around the field  at a rate which is
$a \gamma$ times faster than the rate of rotation of the orbit direction.
At HERA energies $a \gamma$ is between 60 and 68.
So  it can be arranged that small  commuting deflections of the orbit
can result in large noncommuting precessions of the polarisation
vector which can be utilised to rotate the polarisation from the vertical
to the longitudinal direction and vice versa. 
For HERA, the Mini-Rotator design of Buon and Steffen was adopted \cite
{bs85}.
The first pair of these spin rotators was intalled at the East straight section
for the HERMES experiment.

Synchrotron radiation not only generates polarisation
but can also cause depolarisation (Article I). This is especially the
case in the presence of spin rotators. Furthermore the ratio:
(depolarisation rate/polarisation rate) increases strongly with
energy. However, the depolarising effects
can in principle be minimised by special choice of the optic
called `spin matching' \cite{br98}.
But owing to the  difficulty of obtaining reliable
numerical predictions of the polarisation in the presence of rotators
throughout the preparatory stage of the project
and because of the initially very pessimistic predictions, it was by
no means clear that longitudinal polarisation could be obtained
even after spin matching.

Nevertheless, on the first attempt with the rotators 
switched on at the chosen energy of $27.5~GeV$, a longitudinal electron 
polarisation of about $56\%$ was attained.
This is to be compared with the $65\%$ polarisation attained 
immediately beforehand with the rotators turned off.

This was the first time in the history of high energy storage ring physics
that longitudinal polarisation had been attained.
Space limitations prevent my giving more details here but
complete information and diagrams can be found in \cite{ba95}.

Subsequently longitudinal polarisation levels of about $70\%$ for periods
of up to ten hours for positrons in collision with tens of milliamps of 
high energy protons have been achieved. Furthermore by measuring the
 polarisation of 
individual positron bunches the influence of the beam--beam interaction
on the polarisation has been observed and found to be exotic. For example
positron bunches in collision with protons can have a higher polarisation
than non--colliding bunches and the  polarisation of both groups of 
bunches is very sensitive to the tunes of the machine. Obviously, complicated
resonance phenomena are at work \cite{ba98}.

\section*{Outlook}
In the year 2000 two more pairs of spin rotators will be installed so that
three HERA experiments can work with longitudinally polarised electrons
or positrons.

\end{document}